\documentclass[12pt]{article}
\usepackage{graphicx}
\usepackage{amsmath}
\usepackage{amssymb}
\usepackage{times}

\begin{document}
\thispagestyle{empty}
\vspace*{30pt}\begin{center}
{\LARGE\textbf{The structure of cosmic time}}\bigskip\vspace{10pt}\\
\textmd{George Sparling\\Laboratory of Axiomatics\\Department of Mathematics\\
University of Pittsburgh\\Pittsburgh, Pennsylvania, USA}\vspace{10pt}\\
\vspace{50pt}
{\large\textbf{Abstract}}\\\vspace{5pt}
\begin{quote}\textbf{Following the approach of Julien Lesgourgues [1], we analyze the mathematical structure of the time co-ordinate of present day cosmological models, where these models include a cosmological constant term to account for the observed acceleration of the universe [2], [3]: we find that in all cases, except for a set of measure zero in the parameter space, the time is given by an (abelian) integral on a torus; the imaginary period of this integral then gives a natural periodicity in imaginary time for the universe; following Stephen Hawking [4], [5], this periodicity may be interpreted either as giving a fundamental mass scale for the universe, or (using Planck's constant) a fundamental temperature, or both. The precise structure that emerges suggests that the structure of time can be regarded as an order parameter arising perhaps in a phase transition in the early universe; one might hope that this structure would be predictable in any fundamental theory.  }
\end{quote}
\end{center}
\eject\noindent
\section*{Introduction} 
The study of cosmology has been transformed recently by the discovery that the universe is apparently accelerating [2], [3].  If this is true, the standard cosmological models have to be updated to include the acceleration.    Typically these refined models involve the introduction of a cosmological constant term in the formula governing the Hubble function (see below).  It is this case that we analyze here, although much of the structure we discuss would go through with appropriate modifications for other kinds of Hubble functions.   The space-time metrics in question can all be written in the form:
\[ g = ds^2 = c^2 dt^2 - a^2(t) g_\kappa.\]
Here $g_\kappa$ is a three-dimensional metric of constant scalar curvature $\kappa$ (which is naturally compact if $\kappa > 0$ and could be either compact or non-compact, for  the other values of $\kappa$).   Note that the time co-ordinate $t$ is canonical in that $\partial_t$ is of constant size and perpendicular to the surfaces of homogeneity, which are given by the equation $t = $ constant.   The function $a(t) > 0$ is the scale factor for the $t = $ constant three-geometries.  In turn $a(t)$ is related to the Hubble function  $H(t)$ by the differential formula:  $da(t) = a(t) H(t) dt$.    As discussed nicely by  Lesgourgues [1], the Hubble function $H(t)$ is related to the matter and energy content of the universe by the following formula:
\[ 3c^2 H^2(t) = 8\pi \mathcal{G}(\rho_R + \rho_M) - 3\kappa c^4a^{-2}(t)  + \Lambda c^2.\]
Here $\Lambda$ is the cosmological constant term, $c$ is the speed of light and $\mathcal{G}$ is Newton's gravitational constant.   Also $\rho_R > 0$ represents the radiation content of the universe and $\rho_M > 0$ the matter content.    For the purpose of this work, we take as given that these have the standard characteristic decay rates that $\rho = \rho_R a(t)^4$ and $\sigma = \rho_M a^3(t) $ are positive constants.  Then our formula for the the Hubble function can be re-written:
\[ 3c^2 a^4(t) H^2(t) = 8\pi \mathcal{G}(\rho + a(t)\sigma) - 3\kappa c^4 a^2(t)  + \Lambda c^2 a^4(t).\]
The structure of this relation is clarified if we take $a(t)$ as the basic variable, in which case the parameter $t$ is determined implicitly by the differential formula:  $\displaystyle{dt =  \frac{da}{a H(a)}}$, where now the Hubble function is regarded as a function of $a$. In the model given above, $H(a)$ is determined by the formula:
\[ 3c^2 a^4 H^2 = 8\pi \mathcal{G}(\rho + a\sigma) - 3\kappa c^4 a^2  + \Lambda c^2 a^4.\]
Using the scale factor $a$ as co-ordinate, the metric now takes the form:
\[ g = a^2\left(\frac{c^2 da^2}{a^4 H^2(a)} - g_\kappa\right).\]
We find it convenient for our exposition to use instead of $a$, the inverse scale factor $u = a^{-1}$ as our co-ordinate.  Then the metric becomes:
\[ g = u^{-2} \left(\frac{c^2 du^2}{H^2(u)} - g_\kappa\right),\]
\[ 3c^2 H^2 = 8\pi \mathcal{G}(\rho u^4 + u^3\sigma) - 3\kappa c^4 u^2  + \Lambda c^2.\]
Lastly, we absorb a factor of $c^{-1}$ into $H$, so we put $H = cy$, bringing the metric to its final form:
\[ g = u^{-2} \left(\frac{du^2}{y^2(u)} - g_\kappa \right),\]
\[ 3c^4 y^2 = 8\pi \mathcal{G}(\rho u^4 + u^3\sigma) - 3\kappa c^4 u^2  + \Lambda c^2.\]
Here both $y$ and $u$ have the dimensions of inverse length.   Note that, for an expanding universe, $u$ decreases in the direction of increasing time.  We call $y$ the Hubble scale.   It is related to the standard time co-ordinate $t$ by the differential equation: $dt  = - (ucy)^{-1} du$.  Also of interest is the conformal time parameter $\tau$, defined by the relation $d \tau = - (cy)^{-1} du$.  It is the time co-ordinate such that $\partial_\tau$ is a conformal Killing vector field, perpendicular to the surfaces of homogeneity.    Using conformal time, the metric is written $g = u^{-2} (c^2 d\tau^2 - g_\kappa)$, so the metric $c^2 d\tau^2 - g_\kappa$ represents the conformal structure of the physical metric.  We regard this conformal structure as fundamental, defining the basic causal  structure of the universe, the actual metric being given by the conformal factor $u^{-2}$ applied to this conformal metric.
\\\\
We call the right-hand side of the polynomial equation relating $y$ and $u$ the cosmic quartic.   The equation itself is of a standard type, occurring in the basic theory of Riemann surfaces, defining (if $u$ and $y$ are allowed to be complex numbers) a torus in the complex $(u, y)$-plane, provided that the quartic has distinct roots, which is the generic case. It is the analysis of this torus that will occupy us in this work.
\eject\noindent
\begin{itemize}\item  In section one below, for completeness, we develop the connection and curvature of the metric.  Our main purpose here is to emphasize that the Hubble function is co-ordinate independent and directly measurable, in principle.
\item  In section two, we show how to redefine the various variables to formulate the cosmic balance equation, essentially identical to that of Lesgourgues [1], except that we arrange the definitions such that the key quantities are all positive.   Whenever we make variable changes, we relate these directly to the parameters of the balance equation.
\item In section three we give an overview of the structure of the cosmic quartic.   We make rescalings of the various variables involved, so that the quartic is reduced to the convenient form $\displaystyle{x^4 + x^3 - \frac{9R}{32} x^2 + \frac{S}{512}}$.  Here $x = \rho \sigma^{-1} u$.  Also the parameter $R$ is a positive multiple of $ \kappa\rho \sigma^{-2}$ and the parameter $S$ is a positive multiple of $ \Lambda \rho^3 \sigma^{-4}$.   So we have a two-parameter $(R, S)$-space of quartics to examine, with the axis $R = 0$ corresponding to Euclidean topology and the axis $S = 0 $ corresponding to the case of vanishing cosmological constant.
\item  In section four we show how to parametrize the roots in an efficient way and give examples of the various root types that occur. 
\item  In section five, we bring out the invariant structure of the quartic: it is based on using a two-component spinor $\pi_A$, appropriate ratios of whose components determine the parameter $u$.  We find that the conformal structure of the metric takes a natural form:  $(y^{AB} \pi_A d\pi_{B})^2 - G_\kappa$, where $y^{AB}= - y^{BA}$ and $y^{AB}y^{CD} \pi_{A}\pi_{C}H_{BD}^{EF}\pi_E\pi_F = 4$.  Here $H_{AB}^{CD}$ is a constant spinor, which is symmetric in the pairs $AB$ and $CD$ and is trace-free.  The spinor $H_{AB}^{CD}$ incorporates the parameters defining the universe.  The conformal time is then defined by the formula $cd\tau = y^{AB} \pi_A d\pi_{B}$.  The various special cases, where the quartic has repeated roots are treated one by one.
\item In section six, we carefully analyze the general case where the quartic is non-degenerate.   The parameter $\tau$ now has two periods, one real and one purely imaginary.   It is shown how to conveniently make reparametrizations to calculate these periods explicitly.  
\item In section seven, the results are summarized and the structure of the time parameter is discussed.   The measurement problem is analyzed and discussed in relation to present knowledge.
\end{itemize}  
\eject\noindent
\section{General cosmological metrics}
The cosmological structure of the universe is apparently consistent with being conformally flat.   As described in the introduction,  current cosmological models use the following metric, which is spatially homogeneous and isotropic:
\[ g =  u^{-2}\left[ \frac{du\otimes du }{y^2(u)} -  \frac{dr\otimes dr}{1 - \kappa r^2} - r^2 d\theta\otimes d\theta - r^2\sin^2(\theta) d\phi\otimes d\phi )\right].\]
Here $u$ is positive and the Hubble scale $y(u)$ is a smooth positive function, to be determined by the phenomenology. In this section we derive the basic properties of this metric, for general $y(u)$.  We take the domain of the function $y$ to be the open interval $(U_0, U_1)$, where $0 \le  U_0 < U_1 \le \infty$.   Physically $u$ determines the inverse of the radius of curvature of the level surfaces $u = $ constant, so $U_0^{-1}$ and $U_1^{-1}$ would then determine the maximum and minimum possible radii, respectively, if either or both exist.  The co-ordinate $r$ has the range $(0, \infty)$, unless $\kappa > 0$, in which case the range of $r$ is the open interval $(0, \kappa^{-\frac{1}{2}})$.  The co-ordinate $\theta$ has the range $(0, \pi)$ and the co-ordinate $\phi$ has the range $(-\pi, \pi)$.   Also $\kappa $ is an unknown constant, determining the spatial geometry:   
\begin{itemize} \item $\kappa  = 0$ gives  Euclidean spatial sections. 
 \item $\kappa  > 0$ gives spatial sections that  are three-spheres.
 \item $\kappa  < 0$ gives spatial sections that  are hyperbolic three-spaces of constant negative curvature.
\end{itemize}
Experimentally, it appears that the value $\kappa  = 0$ is presently favored. 
By an appropriate constant rescaling the $r$ co-ordinate and the function $y$, if necessary, it is sufficient to restrict to the cases $\kappa  = 1$, $\kappa  = 0$, or $\kappa  = - 1$.  Also, in each case the spatial sections could be quotiented by a discrete symmetry group to give a spatial section with non-trivial fundamental group.
\\\\The metric $g$ is clearly conformally flat, so has vanishing Weyl tensor.  Using Maple, the covariant Ricci tensor of $g$, $R_{ab}$,  takes the following form:
\[ R_{ab} =   (-uyy' + 3y^2 + 2ku^2) g_{ab} -2 T_a T_b ( yuy'  + \kappa u^2).\] 
Here $T = - (uy)^{-1} du $ is a unit timelike future pointing co-vector perpendicular to the surfaces of homogeneity $u  = $ constant.
The Ricci scalar $R$ is:
\[   R = 6(-uyy' + 2y^2 + \kappa u^2).\] 
\eject\noindent
The Einstein tensor $G_{ab}$ is then:
\[ G_{ab}   =  R_{ab} - \frac{R}{2} g_{ab} = (2uyy' - 3y^2 - \kappa u^2) g_{ab}-2 T_a T_b( yuy'  + \kappa u^2) .\]
\begin{itemize} \item The metric is flat if and only if $G_{ab}$ vanishes,  which gives the relation: \[ y^2 = - \kappa u^2.\] This entails that $\kappa  < 0$.  Then $y = u \sqrt{-\kappa }$ and is defined for all positive $u$.
\item 
The metric is of constant curvature, if and only if $G_{ab}$ is proportional to $g_{ab}$,  which gives the relation: 
\[ y^2 = C - \kappa u^2.\] 
Here  $C$ is an integration constant.   When $y^2 = C - \kappa u^2$,  the Einstein tensor reduces to $G_{ab} = - 3Cg_{ab}$.
\begin{itemize} \item If $\kappa  =0$, we need $C > 0$ and then $y$ is defined for all positive real $u$.
\item If $\kappa  > 0$, we need $C > 0$ and then $y$ is defined for the $u$-interval $(0, \sqrt{C\kappa^{-1}})$.
\item If $\kappa  < 0$ and $C \ge 0$, then $y$ is defined for the $u$-interval $(0, \infty)$.
\item If $\kappa  < 0$ and $C <  0$, then $y$ is defined for the $u$-interval $(\sqrt{C\kappa^{-1}}, \infty)$.
\end{itemize} 

\end{itemize} Henceforth we assume that the metric is not of constant curvature on any open set, so that the quantity $yuy'  + \kappa u^2$ is never zero on any open set.
\begin{itemize} 
\item Taking the divergence of both sides of the equation for the Einstein tensor, we get, using the fact that $\partial_a u = (du)_a = - uy T_a$ and the fact that the Einstein tensor has zero divergence:
\[ 0 = - 2( yuy'  + \kappa u^2) (T_b\partial^a T_a  + T_a \partial^a T_b)  + 3 uy T_b(\kappa u^{2} + y^2)'  \]
\[ = - 2( yuy'  + \kappa u^2) (T_b\partial^a T_a  + T_a \partial^a T_b - 3yT_b), \]
\[ 0 = T_b\partial^a T_a  + T_a \partial^a T_b - 3yT_b.\]
Taking the trace of this equation with $T^b$, we get the relations:
\[\partial^a T_a = 3y,  \]
\[ T_a \partial^a T_b =0.\]
\end{itemize} 
This tells us that $T^a$ is a parallelly propagated unit geodesic timelike vector field. Also measurement of the quantity $\frac{1}{3}\partial_a T^a$, one third of the covariant  divergence of the vector field, $T^a$, determines the Hubble scale factor $y$. 
\eject\noindent
\section{ The cosmic quartic and the balance equation} 
The specific metric discussed in cosmology takes the form:
\[ ds^2 =   u^{-2} \left[ \frac{ 3c^4 du^2}{8\pi\mathcal{G}(\rho u^4 + \sigma u^3) - 3\kappa c^4u^2+  \Lambda c^2 } -  \frac{dr^2}{1 - \kappa r^2} - r^2 (d\theta^2 + \sin^2(\theta) d\phi^2)\right] \]
So we have the following formula for the Hubble scale $y$:
\[ 3c^4 y^2 = 8\pi\mathcal{G}(\rho u^4 + \sigma u^3) - 3\kappa c^4u^2+  \Lambda c^2 .\]
Here $c > 0$ is the speed of light, $\mathcal{G} > 0$ is Newton's gravitational constant,  $\rho >0$ measures the radiation density and $\sigma > 0$ measures the matter density.   Finally $\Lambda$ is the cosmological constant term (which, if positive, dominates the geometry when $u$ is small).  For the remainder of this section we will assume the generic case that $\Lambda \ne 0$.
We can rewrite the equation for $y$ in terms of present day quantities:
\[ 3c^4 y^2 = 8\pi\mathcal{G}\rho_0\left( \frac{u}{u_0}\right)^4 + 8\pi\mathcal{G}\sigma_0 \left(\frac{u}{u_0}\right)^3 - 3 \kappa c^4u_0^2\left(\frac{u}{u_0}\right)^2+  \Lambda c^2 \]
\[   Z^2=  F(X), \hspace{10pt} F(X) = \Omega_R X^4 + \Omega_M X^3 -\kappa \Omega_\kappa  X^2+  \epsilon\Omega_\Lambda. \]
Here $\kappa ^3 = \kappa $,  $\epsilon^2 = 1$, $u_0$ is the present value of $u$ and $y_0$ is the present day value of $y$.  Also we have:
\[ Z = \frac{y}{y_0}, \hspace{10pt} X = \frac{u}{u_0}, \hspace{10pt} y_0^2 =  \frac{|\Lambda|}{3c^2\Omega_\Lambda},  \hspace{10pt} u_0^2 = \frac{|\Lambda|\Omega_\kappa }{3c^2\Omega_\Lambda}, \]
\[ \rho= \rho_0u_0^{-4} = \frac{3y_0^2c^4\Omega_R}{8\pi\mathcal{G}u_0^3} = \frac{9c^6}{8\pi\mathcal{G}}\frac{\Omega_R\Omega_\Lambda}{|\Lambda|\Omega_\kappa ^2},  \hspace{10pt} \sigma = \sigma_0u_0^{-3} = \frac{3y_0^2c^4\Omega_M}{8\pi\mathcal{G}u_0^3} = \frac{3\sqrt{3}c^5}{8\pi\mathcal{G}}\frac{\Omega_M\Omega_\Lambda^{\frac{1}{2}}}{|\Lambda|^{\frac{1}{2}}\Omega_\kappa ^{\frac{3}{2}}} .  \]
\[ \Omega_R = \frac{8\pi\mathcal{G}\rho_0}{3y_0^2c^4} , \hspace{10pt} \Omega_M = \frac{8\pi\mathcal{G}\sigma_0}{3y_0^2c^4} , \hspace{10pt} \Omega_\kappa  = \frac{u_0^2}{y_0^2} , \hspace{10pt} \Omega_\Lambda = \frac{|\Lambda|}{3c^2y_0^2}, \]
\[ 1 =  \Omega_R + \Omega_M - \kappa \Omega_\kappa  +  \epsilon \Omega_\Lambda, \]
This last equation is called the cosmic balance equation, reflecting the relative contributions of matter ($\Omega_M$), radiation ($\Omega_R$), topology ($\Omega_\kappa$) and the cosmological constant ($\Omega_\Lambda$).  The quantities $Z, X, \Omega_R, \Omega_M, \Omega_\Lambda$ and $\Omega_\kappa $ are all dimensionless and positive. 
\section{A unified treatment of the cosmic quartic}
Consider the equation for the Hubble parameter:
 \[ 3c^4 y^2 = 8\pi\mathcal{G}(\rho u^4 + \sigma u^3) - 3\kappa c^4u^2+  \Lambda c^2 .\]
 Make a change of variables:
 \begin{itemize} \item    $\displaystyle{u = \frac{x \sigma}{\rho}}$,
 \item $\displaystyle{y =Y \sqrt{8\pi\mathcal{G}\sigma^4\rho^{-3}c^{-2}}}$,
 \item $\displaystyle{R =  \frac{4\kappa c^4\rho}{3\pi\mathcal{G}\sigma^2}}$, 
 \item $\displaystyle{S = \frac{64\Lambda c^2\rho^3}{\pi\mathcal{G}\sigma^4}}$.
 \end{itemize} 
 Note that $R$ and $S$ are positive multiples of $\kappa $ and $\Lambda$, respectively and $x$, $Y$, $R$ and $S$ are pure numbers (dimensionless).  When $\kappa $ and $\Lambda$ are non-zero, we have:
\begin{itemize} \item $ \displaystyle{u = \frac{x|\Lambda|^{\frac{1}{2}}}{c\sqrt{3}}\left(\frac{\Omega_M\Omega_\kappa ^{\frac{1}{2}}}{\Omega_R\Omega_\Lambda^{\frac{1}{2}}}\right)}$,
\item $ \displaystyle{ y = Y|\Lambda|^{\frac{1}{2}} \left( \frac{\Omega_M^2}{\Omega_R^{\frac{3}{2}}\Omega_\Lambda^{\frac{1}{2}}}\right)}$, 
\item $ \displaystyle{ R = \frac{32\kappa}{9}\left( \frac{ \Omega_R \Omega_\kappa }{\Omega_M^2}\right)}$, 
\item $ \displaystyle{S = \frac{512\Lambda}{|\Lambda|}  \left(\frac{\Omega_R^3\Omega_\Lambda }{\Omega_M^4}\right)}$.
\end{itemize} 
In terms of these variables the Hubble parameter equation becomes:
 \[ 3Y^2 = H(x), \hspace{10pt} H(x) = x^4 + x^3 -\frac{9R}{32}x^2+  \frac{S}{512}.\] 
From the present viewpoint, to precisely determine the metric of space-time, the very difficult experimental task is to measure accurately the pure numbers $R$ and $S$.  \eject\noindent Note that a priori the parameters $R$ and $S$ are continuous variables.  In particular, if the sign of $R$ is determined, then we will know the topological type of the spatial geometry.   Currently it seems to be believed that the universe is spatially Euclidean.   However it is rather  dangerous to take this seriously, as this requires that $R$ be exactly zero.  A small deviation in $R$ from zero, of either sign would drastically alter the spatial character of the metric, without substantially altering its temporal evolution on time-scales comparable with the present age of the universe.  In this section, we give a unified treatment of the cosmic quartic $H(x)$, with no prior assumptions about the values of the quantities $R$ and $S$.  First consider the behaviour of the roots of $H$ as the parameters $R$ and $S$ are varied.  The resultant $J(R, S)$ of $H$ and its derivative with respect to $x$ factors as: 
\[ J(R, S) = 2^{-25} S(64S^2-432S(3R^2+12R + 8) +729R^3(9R+8R^3)).\]
Solving the equation $J(R, S) = 0$, for $S$ in terms of $R$, gives three roots, $\{ 0, S_+, S_-\}$, where $S_\pm$ are given as follows:
\[  S_\pm = \frac{9}{8}\left((3R+6-2\sqrt{3})(3R+6+2\sqrt{3}) \pm 24(R + 1)^{\frac{3}{2}}\right).\]
Deleting the zero set of $J$ divides its complement in the $(R, S)$ plane into five connected open regions, where the character of the roots is the same in each connected component. 
We discuss each  in turn:
\begin{itemize} \item Region I: $0 < S < S_-, R > 0$.  \\In this region $H$ has four real distinct roots, two positive and two negative. 
\item Region II: $S_- < S < S_+$ and $S > 0$. \\In this region $H$ has two real distinct roots, both negative and two non-real roots.
\item Region III: $S > S_+$ and $S > 0$.\\ In this region $H$ has all its roots non-real.
\item Region IV: $S < 0$ with $R < -1$ or $R > 0 $;  or $S_+ < S < 0$, with $-1 \le R < - \frac{8}{9}$; or $S < S_-$, with $-1 \le R \le 0$. \\In this region $H$ has one real root of each sign and two non-real roots.
\item Region V: $S_- < S < 0$ and $- 1 < R < 0$.  \\ In this region $H$ has one positive real root  and three distinct negative roots.  This is the only bounded region.
\end{itemize} 
The edges of the regions correspond to cases that the roots are repeated.
\begin{itemize} \item The origin $R = S = 0$ is a boundary point of all regions except for region III.  The roots are $0, 0, 0$ and $-1$.  This is the Euclidean case with $\Lambda = 0$.
\item All the regions except for region I have a common boundary point with $R = - \frac{8}{9}$ and $S = 0$.   The roots in this case are $-\frac{1}{2}$ and $0$, each double roots.
\item Regions IV and V have a boundary cusp point at $R = - 1$ and $S = - \frac{27}{8}$.   The roots in this case are a simple root of $\frac{1}{8}$ and a triple root of $-\frac{3}{8}$.
\end{itemize}
Deleting these three special points, the boundaries split into seven open connected curve components:
\begin{itemize} 
\item I-II: The curve separating regions I and II is given by the equation $S = S_-$, with $R > 0$.  Here the roots are all real with two equal positive roots and two unequal negative roots.
\item II-III: The curve separating regions II and III is given by the equation $S = S_+$, with $R > - \frac{8}{9}$.  Here the repeated  roots are negative and  there are two non-real roots.
\item III-IV: The curve separating regions III and IV is given by the equation $S = 0 $, with $R < - \frac{8}{9}$.  Here the repeated  roots are zero  and  there are two non-real roots.
\item IV-VA: The first curve separating regions IV and V  is given by the equation $S = S_+ $, with $- 1< R < - \frac{8}{9}$.  Here the repeated  roots are negative  and  there are two other real roots one of each sign.   The repeated roots are the most negative.
\item IV-VB: The second curve separating regions IV and V  is given by the equation $S = S_- $, with $- 1< R < 0$.  Here the repeated  roots are negative  and  there are two other real roots one of each sign.   The non-repeated negative root is the most negative.
\item II-V: The curve separating regions II and V is given by the equation $S = 0 $, with $ - \frac{8}{9} < R < 0$.  Here the repeated  roots are zero  and  there are two other real roots, both negative.
\item I-IV: The curve separating regions I and IV is given by the equation $S = 0 $, with $R > 0$.  Here the repeated  roots are zero and there are two other real roots, one of each sign.
\end{itemize}
\eject\noindent
We now summarize the structure of the quartic according to the signs of $\Lambda$ and $\kappa $.    We first look at the cases where one or both of these quantities vanishes.
\begin{itemize} \item In the case that $\Lambda = 0$, we have $S = 0$. 
Then we have: 
\[ H(x) = x^2\left(x^2 + x - \frac{9R}{32}\right), \hspace{4pt} \textrm{with roots} \hspace{4pt} \left\{ 0, 0, \frac{1}{8} (- 4 \pm \sqrt{18R + 16})\right\}\]
\begin{itemize} \item   The Euclidean point is then $R = 0$ and $H(x) = x^3(x  + 1)$.
\item 
The elliptic case is $R > 0$, so is the I-IV boundary.    Then $H(x) $ has a positive and a negative real root.
\item The hyperbolic case has $R < 0$, so has three parts: 
\begin{itemize}\item the II-V boundary, with $ - \frac{8}{9} < R < 0$, so $H(x) $ has two  negative real roots.
\item  the special point $R = - \frac{8}{9}$, when $H(x) = \frac{1}{4}x^2(2x + 1)^2$.
\item The III-IV boundary, with $R < - \frac{8}{9} $, so $H(x) $ has two non-real roots.
\end{itemize} 
\end{itemize}
 \item The Euclidean case with $\Lambda > 0$ has $R = 0$ and breaks up into three subcases:
\begin{itemize} \item If $0 < S < 54$, then we are in the interior of region II.
\item If $S> 54$, then we are in the interior of region III.
\item If $S = 54$, then we lie on the II-III boundary curve.  In this case we have:
\[ H(x) = x^4 + x^3 + \frac{27}{256} = \frac{1}{256} (16x^2-8x+3)(4x+3)^2\]\[  = \frac{1}{256}(4x + 3)^2(4x-1+i \sqrt{2})(4x-1-i \sqrt{2}).\]
\end{itemize} 
 \item The Euclidean case with $\Lambda < 0$ has $R = 0$.   The quartic then lies in region IV, so has one real root of each sign and two non-real roots. 
\end{itemize} 
\eject\noindent
Finally we summarize the various generic cases where neither $\Lambda$ nor $\kappa $ vanishes.
\begin{itemize} 
\item The elliptic case with $\Lambda > 0$ corresponds to the open first quadrant of the $(R, S)$-plane.   Here we have all of region I and the parts of the I-II boundary and of region II in the first quadrant.
\item The elliptic case with $\Lambda < 0$ corresponds to the open fourth quadrant of the $(R, S)$-plane.   Here we have the parts of region IV in the fourth quadrant.
\item The hyperbolic case with $\Lambda > 0$ corresponds to the open second quadrant of the $(R, S)$-plane.   Here we have the parts of regions II and III in the second quadrant and the parts of the II-III boundary curve in the open second quadrant.
\item The hyperbolic case with $\Lambda < 0$ corresponds to the open third quadrant of the $(R, S)$-plane.   Here we have all of region V, both boundary curves IV-VA and IV-VB, the cusp point $R = - 1$, $S = - \frac{27}{8}$ and the part of region IV open fourth quadrant.
\end{itemize} 
\eject\noindent
\section{Parametrizing the roots of the cosmic quartic}
For many purposes it is desirable to have good control over the roots of the quartic.   One method to do this is as follows.   Given $R$ and $S$, introduce a new parameter $P\ne 0$ by the implicit formula:
\[ S = \frac{1}{8P}(P-1)(9R-8P+8)^2.\]
Multiplying out, this gives a cubic equation for $P$, which is straightforward to solve for $P$ in terms of $R$ and $S$ by standard methods:
\[ 0 = g(P) = 64P^3- 48P^2(3R + 4)+P(3(3R + 8)(9R + 8)-8S)- (9R + 8)^2.\]
Note that for given $R$ and $S$, we have $g(0) = - (9R + 8)^2$ and $g(1) = - 8S$, so, if  if $S \ge 0$,  the equation $g(P) = 0$ has a solution $P \ge 1$, whereas if $S < 0$,  there is a solution in the interval $0 < P < 1$, unless $R = - \frac{8}{9}$.   
\begin{itemize} \item   In the exceptional case that $R = - \frac{8}{9}$, we have a direct factorization of the quartic:
\[ H(x) = x^4 + x^3 + \frac{1}{4}x^2 + \frac{S}{512} = \frac{1}{1024}(32x^2 + 16x + i\sqrt{2S})(32x^2 + 16x - i\sqrt{2S}).\]
The roots are $\displaystyle{\{ \frac{1}{8}(- 2 \pm \sqrt{4 + 2i\sqrt{2S}}), \frac{1}{8}(- 2 \pm \sqrt{4 - 2i\sqrt{2S}})\}}$. 
\begin{itemize} \item  If $S> 0$ all roots are non-real, with equal sizes for their imaginary parts and with two positive real parts two negative.
\item   If $S = 0$, the roots are $0$ and $-\frac{1}{2}$, each a double root. 
\item  If  $- 2 < S< 0$, then all four roots are real, with  three negative and one positive.
\item If $S =- 2$, then there is a double root of $ - \frac{1}{4}$, a third more negative root $- \frac{1}{4}(\sqrt{2} + 1) $ and a positive root $\frac{1}{4}( \sqrt{2} - 1) $.
\item If $S < -2$, there are two real roots, one of each sign, and two non-real roots each with real part $- \frac{1}{4}$.
\end{itemize} 
\end{itemize} 
 Henceforth we assume the generic case that $R \ne -\frac{8}{9}$.    The cubic for $P$ simplifies if we write $S = \frac{27}{8}(T - R^2)$, where $T$ is real.   Then the critical points of $g(P)$ are $\frac{1}{8}(6R + 8 \pm 3\sqrt{T})$.
 \begin{itemize}\item If $S < - \frac{27}{8}R^2$, then $T  < 0$, so $g(P)$ has no real critical points. Then $g(P)$ is everywhere increasing and its real root is unique, is larger than $1$, and varies smoothly with $R$ and $S$.
\item If $S = - \frac{27}{8}R^2$, then $T = 0$ and $g(P)$ has one critical point only, which is also its inflection point, so $g(P)$ is everywhere increasing, so its real root is unique.  Explicitly, we have, for this case:
\[ P = 1 + \frac{3R}{4}(1 - (1 +R^{-1})^{\frac{1}{3}}).\]
So for $R> 0$, $P$ lies in the range $\frac{3}{4} < P < 1$ and decreases towards the value $\frac{3}{4}$, as $R$ gets large.
\item If $S > - \frac{27}{8}R^2$, then $T > R^2$.  Then the value of $g(P)$ at the largest critical point $P = \frac{1}{8}(6R + 8 + 3\sqrt{T})$ is $\frac{27}{4}(R-\sqrt{T})(T + 4(R + 1)(R + \sqrt{T}))$.  Since $ T> R^2$, this value is plainly negative.  So in this case $g(P)$ has a unique real root, which is larger than $\frac{1}{24}(18R + 24 + \sqrt{81R^2 + 24S})$ and which varies smoothly with $R$ and $S$.
\end{itemize}
The resultant of $g(P)$ with its derivative factors as $-2^{42}J(R, S)$, so the roots of $g(P)$ change character when those of $H(x)$ do.
\begin{itemize} 
\item At  $(0 , 0)$ in the $(R, S)$-plane, all three roots are $1$.
\item At  $(-\frac{8}{9} , 0)$ in the $(R, S)$-plane, one root is $1$ and the other root is repeated and is $P =0$.
\item At  $(-1 ,- \frac{27}{8})$ in the $(R, S)$-plane, all three roots are $P = \frac{1}{4}$.
 \item In region I, there are three real roots for $P$, all positive and larger than $1$.
\item  In region II, there is one real root for $P$, which is positive and larger than $1$.
\item  In region III, there are three real roots for $P$, with exactly one positive;  this root is larger than $1$; except along the vertical line $R = - \frac{8}{9}$, when one root is zero, one is positive and larger than one and one is negative.
\item  In region IV, there is exactly one real root for $P$, which is positive and smaller than $1$,  except along the vertical line $R = - \frac{8}{9}$, when one root is zero and the other two are non-real with real part $\frac{1}{2}$.
\item In region V, there are three positive real roots for $P$, all smaller than $1$, except along the vertical line $R = - \frac{8}{9}$, when one root is zero and the other two are real and smaller than $1$.
\item On the I-II boundary curve there are two equal real roots and a third larger real root, all positive.
\item On the II-III boundary curve there are two equal negative real  roots and a third positive real root, which is larger than $1$.
\item On the III-IV boundary curve there are two equal negative real  roots $P = \frac{1}{8}(9R + 8)$, with $R < - \frac{8}{9}$ and the third root is $P = 1$.
\item On the boundary curve IV-VA, all three roots are positive and smaller than $1$, with the two smallest equal.
\item On the boundary curve IV-VB, all three roots are positive and smaller than $1$, with the two largest equal; except at the point $(-\frac{8}{9}, -2)$ of this curve, when one root is $P = 0$ and the other two are each $P = \frac{1}{2}$.
\item On the II-V boundary curve there are two equal positive real  roots $P = \frac{1}{8}(9R + 8)$, with $- \frac{8}{9} < R < 0$, both smaller than $1$ and the third root is $P = 1$.
\item On the I-IV boundary curve there are two equal positive real  roots $P = \frac{1}{8}(9R + 8)$, with $R > 0$, both larger than $1$ and the third root is $P = 1$.
\end{itemize} 
Replacing $S$ by its expression in terms of $P$ and $R$, the cosmic quartic factorizes readily:
\[ H(x) = x^4 + x^3 - \frac{9R}{32}x^2 + \frac{S}{512} = \frac{Q_+Q_-}{4096}, \]
\[ Q_\pm = 64x^2 + 32x - 9R - 8  + 8P \pm (\sqrt{P})^{-1}(8(4x + 1)P - 9R - 8).\]

The roots are then:
\[  \frac{1}{8}\left( - 2-2\sqrt{P} \pm \sqrt{(1 + (\sqrt{P})^{-1})(3\sqrt{R+1}+1-2\sqrt{P})(3\sqrt{R+1}-1+2\sqrt{P})}\right),\]
\[  \frac{1}{8}\left(  - 2+2\sqrt{P}\pm \sqrt{(1 - (\sqrt{P})^{-1})(3\sqrt{R+1}+1+2\sqrt{P})(3\sqrt{R+1}-1-2\sqrt{P})}\right).\]
\eject\noindent
We give at least one example for each case:
\begin{itemize} \item At $(0, 0)$, $H(x) = x^4 + x^3 = x^3(x + 1)$, with roots $\{ 0, -1\}$.
\item At $(-\frac{8}{9} , 0)$, $H(x) = \frac{1}{4} x^2(2x + 1)^2$, with roots $\{0, -\frac{1}{2}\}$.
\item At $(-1 , \frac{27}{8})$, $H(x) = 2^{-14}(128x^2+192x-9)(128x^2-64x-3)$, with roots $\{\frac{1}{8}, -\frac{3}{8}\}$.
\item At the point $(3, \frac{27}{32})$ in region I, $H(x) = 2^{-14}(128x^2+192x-9)(128x^2-64x-3)$, with roots $\{\frac{1}{16}(4 \pm \sqrt{22}), \frac{3}{16}( -4\pm3\sqrt{2})\}$.
\item At the point $(3, \frac{1369}{9})$ in region II, \\$H(x) = 2^{-9}3^{-2}(96x^2-96x+37)(48x^2+96x+37)$, \\with roots $\{\frac{1}{12}(-12\pm \sqrt{33}), \frac{1}{24}(12\pm i \sqrt{78})\}$.
\}.
\item At the point $(0, \frac{125}{18})$ in region II,\\ $H(x) = 2^{-10} 3^{-2}(96x^2+ 120x+25)(96x^2- 24x+25)$, \\with roots $\{\frac{5}{24}(-3\pm\sqrt{3}), \frac{1}{24}(3 \pm i \sqrt{21})\}$.
\item At the point $(-\frac{3}{4}, \frac{85697}{41472})$ in region II, \\$H(x) = 2^{-18}3^{-4}(4608x^2+4896x+1207)(4608x^2-288x+71)$,\\ with roots $\{\frac{1}{96}(3\pm i\sqrt{133}), \frac{1}{96}(-51 \pm i \sqrt{187})\}$.
\item At the point $(3, 3267)$ in region III,\\ $H(x) = 2^{-9}(16x^2-32x+33)(32x^2+96x+99)$,\\ with roots $\{\frac{1}{8}(-6+\pm 3i \sqrt{6}), \frac{1}{4}(4+\pm i \sqrt{17})\}$.
\}.
\item At the point $(0, \frac{3375}{2})$ in region III, \\$H(x) = 2^{-10}(32x^2-48x+45)(32x^2+80x+75)$, \\with roots $\{\frac{5}{8}(-2+\pm i \sqrt{2}), \frac{3}{8}(2+\pm i \sqrt{6})\}$.
\item At the point $(-\frac{3}{4}, \frac{45387}{512})$ in region III, \\$H(x) = 2^{-18}(512x^2-256x+123)(512x^2+768x+369)$,\\ with roots $\{\frac{3}{32}(-8\pm 3i\sqrt{2}), \frac{1}{32}(8 \pm i \sqrt{182})\}$.
\item At the point $(3, -\frac{3267}{8})$ in region IV, \\$H(x) = 2^{-12}(64x^2+ 16x+33)(64x^2+48x-99)$, \\with roots $\{\frac{3}{8}(-1\pm 2 \sqrt{3}), \frac{1}{8}(-1\pm 4 i \sqrt{2})\}$.
\item At the point $(0, -\frac{27}{2})$ in region IV, \\$H(x) = 2^{-10}(32x^2+ 8x+3)(32x^2+24x-9)$, \\with roots $\{\frac{3}{8}(-1\pm\sqrt{3}), \frac{1}{8}(-1\pm  i \sqrt{5})\}$.
\item At the point $(-1, -\frac{135}{32})$ in region IV, \\$H(x) = 2^{-14}(128x^2+ 48x-9)(128x^2+80x+ 15)$, \\with roots $\{\frac{3}{16}(-1\pm  \sqrt{3}), \frac{1}{16}(-5\pm  i \sqrt{5})\}$.
\item At the point $(-\frac{8}{9}, -\frac{3}{2})$ in region V,\\ $H(x) = 2^{-10}(32x^2+ 24x+3)(32x^2+8x- 1)$,\\ with roots $\{\frac{1}{8}(-1\pm  \sqrt{3}), \frac{1}{8}(-3\pm   \sqrt{3})\}$.
\item At the point $(3, \frac{189}{8})$ on the I-II boundary curve, \\$H(x) = 2^{-12}(64x^2+ 112x+21)(8x - 3)^2$, with roots $\{\frac{3}{8}, \frac{1}{8}(-7\pm  2 \sqrt{7})\}$.
\item At the point $(3, \frac{3645}{8})$ on the II-III boundary curve, \\$H(x) = 2^{-12}(64x^2- 80x+45)(8x + 9)^2$, with roots $\{- \frac{9}{8}, \frac{1}{8}(5\pm  2i \sqrt{5})\}$.
\item At the point $(0, 54)$ on the II-III boundary curve, \\$H(x) = 2^{-8}(16x^2- 8x+3)(4x + 3)^2$, with roots $\{- \frac{3}{4}, \frac{1}{4}(1\pm  i \sqrt{2})\}$.
\item At the point $(-\frac{3}{4}, \frac{729}{128})$ on the II-III boundary curve, \\$H(x) = 2^{-16}(256x^2- 32x+9)(16x + 9)^2$, with roots $\{- \frac{9}{16}, \frac{1}{16}(1\pm  2i \sqrt{2})\}$.
\item At the point $(-\frac{15}{16}, -\frac{3375}{2048})$ on the IV-VA boundary curve,\\ $H(x) = 2^{-20}(32x + 5)(32x - 3)(32x + 15)^2$, with roots $\{- \frac{15}{32},- \frac{5}{32}, \frac{3}{32})\}$.
\item At the point $(-\frac{15}{16}, -\frac{5103}{2048})$ on the IV-VB boundary curve,\\ $H(x) = 2^{-20}(1024x^2 + 448x- 63)(32x + 9)^2$, \\with roots $\{- \frac{9}{32}, \frac{1}{32}(-7\pm  4 \sqrt{7})\}$.
\item At the point $(3, 0)$ on the I-IV boundary curve,\\ $H(x) = 2^{-5}x^2(32x^2 +32x - 27)$, with roots $\{0, \frac{1}{8}(-4\pm   \sqrt{70})\}$.
\item At the point $(-\frac{7}{16}, 0)$ on the II-IV boundary curve, \\$H(x) = 2^{-9}x^2(512x^2 + 512x + 63)$, with roots $\{0, \frac{1}{32}(-16\pm   \sqrt{130})\}$.
\item At the point $(-1, 0)$ on the III-IV boundary curve, \\$H(x) = 2^{-5}x^2(32x^2 + 32x + 9)$, with roots $\{0, \frac{1}{8}(-4\pm  i \sqrt{2})\}$.
\end{itemize}
\eject\noindent
\section{The invariant approach to the cosmic quartic}
We introduce a (real) two-component spinor $\pi_A$, the ratio of whose components represent the variable $x$.     Let $H_{AB}^{\hspace{12pt} CD}$ be a fixed spinor, obeying the following symmetry conditions:
\[ H_{AB}^{\hspace{12pt}CD} = H_{(AB)}^{\hspace{17pt}CD} = H_{AB}^{\hspace{10pt}(CD)}, \hspace{10pt} H_{AB}^{\hspace{12pt} AC} = 0.\]
Introduce a skew spinor symplectic form $y_{AB} = - y_{BA}$ and its (skew) inverse $y^{AB}$, defined in terms of $ H_{AB}^{\hspace{12pt} CD} $ and $\pi_E$ by the relations:
\[ y_{AB} y_{CD} = \pi_{[C}\pi_{[A}H_{B]D]}^{\hspace{17pt}EF}\pi_E\pi_F, \hspace{10pt} y_{AB}y^{AB} = 2,\]
\[ y^{AB}y^{CD} \pi_{A}\pi_{C}H_{BD}^{\hspace{14pt}EF}\pi_E\pi_F = 4.\]
Note that $y_{AB} (k\pi_C) = k^{2} y_{AB}(\pi_C)$ and $y^{AB} (k\pi_C) = k^{-2} y^{AB}(\pi_C)$.
Then the conformal metric $G$ of cosmology achieves the following canonical form:
\[ G =   (y^{AB}\pi_Ad\pi_B)^2 - \frac{dr^2}{1 - \kappa r^2} - r^2 (d\theta^2 + \sin^2(\theta) d\phi^2), \]
Here all the variables $ \pi_A$, $r$, $\theta$,  $\phi$, $y^{AB}$ and $ H_{AB}^{\hspace{12pt} CD}$ are dimensionless.  Also $\kappa^3 = \kappa$.   The information of the cosmic quartic is encoded in the constant spinor $H_{AB}^{\hspace{12pt}CD}$.  In turn the spinor $H_{AB}^{\hspace{12pt}CD}$ may be regarded as tensor $H_a^{\hspace{4pt} b}$, where the Latin indices $a \leftrightarrow (AB)$ are three-dimensional, such that $H_a^{\hspace{4pt} a} = 0$ and $H_{ab} = H_{(ab)}$, so $H_a^{\hspace{4pt}b}$ has five degrees of freedom.
The metric $G$ is invariant under the natural action of $\textrm{GL}(2, \mathbb{R})$ on all the spinors.  Since the spinor $H_{AB}^{\hspace{12pt}CD}$ has equal numbers of upper and lower indices, the action of $\textrm{GL}(2, \mathbb{R})$ factors through the action of $\textrm{SL}(2, \mathbb{R})$ and even of $\textrm{PL}(2, \mathbb{R})$, or the pseudo-orthogonal group, $\textrm{SO}(1, 2)$.  The group action has three degrees of freedom, giving the orbit structure two degrees of freedom.   With respect to a suitable basis, the matrices $G_0$ for the covariant  $\textrm{SO}(1, 2)$ metric and $H$ for the cosmic quartic take the following forms:
\[ G_0  = \hspace{4pt}  \begin{array}{|ccc|}0&0&1\\0&- 2&0\\1&0 &0\end{array}\hspace{3pt},\]
\[ H = \hspace{4pt}  \begin{array}{|ccc|}- \frac{3}{64}R&-1/2&1\\ 0&\frac{3}{32}R&\frac{1}{4}\\\frac{S}{512}&0&- \frac{3}{64} R\end{array}\hspace{3pt}.\]
It is easily checked that $H$ is trace-free and $G_0H$ is symmetric, as required.  
\eject\noindent In this language, a spinor $\sigma^A$ with components $(p, q)$ is encoded in a null vector $s^a$ with components $(p^2, pq, q^2)$.  Also a co-spinor $\pi_A$ with components $(x, y)$ is encoded in a null co-vector $p_a$ with components $(x^2, 2xy, y^2)$. Note that we have $s^ap_a = (\sigma^A\pi_A)^2$.
The characteristic polynomial $p(\lambda)$ of the matrix $H$ is:
\[ p(\lambda) = \lambda^3 -\frac{\lambda}{4096}(8S + 27R^2)- \frac{1}{131072}(32S + 24RS + 27R^3).\]
The condition that the roots be distinct is precisely the vanishing of the polynomial $J(R, S)$.    Also in this language, we have $\pi_A$ proportional to the co-spinors $(x, 1)$ and $(u\rho, \sigma)$.  So if we take as basis spinors, the spinors $\alpha^A = (1, 0)$  and $\beta^A = (0, 1)$, then we have:
\[ \frac{\pi_B\alpha^B}{\pi_A \beta^A} = x =  \frac{\rho u }{\sigma}.\]
The (dimensionless) conformal time parameter is defined by the formula:
\[ d\tau = y^{CD} \pi_{C} d\pi_D.\]
The (dimensionless)  time parameter is defined by the formula:
\[ dt = x^{-1} d\tau =  \frac{\pi_A\alpha^A}{\pi_B \beta^B}y^{CD} \pi_{C} d\pi_D.\]
The philosophy now is that the equation $ y^{AB}y^{CD} \pi_{A}\pi_{C}H_{BD}^{\hspace{14pt}EF}\pi_E\pi_F = 4$ generically defines a torus (i.e. provided the tensor $H_a^b$ has distinct eigen-values), when $y^{AB} $ and $\pi_A$ are allowed to be complex.   Then there are two independent integrals defining each of the quantities $t$ and $\tau$ (i.e. four integrals in all), which can be chosen so that, for each quantity, one of the integrals is real and the other complex.   The periodicity in the variable $\tau$ is the natural geometric periodicity and characterizes the complexified conformal geometry; for example all the Green's functions of solutions of conformally invariant hyperbolic wave equations, such as Maxwell's equations, on the cosmological background will exhibit this periodicity.  Then the corresponding integrals for the co-ordinate $t$ will give the lifetime of the universe for the real integral (if it is finite) and a natural periodicity in imaginary time for the imaginary integral.  Note that the lifetime is always finite in the past going to the big bang as $u \rightarrow \infty$.  
\eject\noindent
Temporarily going back to our initial variables, we have:
\[ t =    - \int \frac{\sqrt{3} cdu}{u\sqrt{8\pi\mathcal{G}(\rho u^4 + \sigma u^3) - 3\kappa c^4u^2+  \Lambda c^2 } }\]
\[ \tau =  -  \int \frac{\sqrt{3} cdu}{\sqrt{8\pi\mathcal{G}(\rho u^4 + \sigma u^3) - 3\kappa c^4u^2+  \Lambda c^2 } }\]
Note that if $\tau$ is used as a parameter and if then the variable $u$ is known as a function of $\tau$, the integral for $t$ simplifies to:
\[ t = \int \frac{d\tau}{u(\tau)}.\]
Passing to the dimensionless versions of these relations, the quartic is now written in terms of the homogeneous variables $x$ and $y$ as:
\[ H(x, y) =  x^4+x^3 y-\frac{9}{32}x^2y^2R+\frac{1}{512}y^4S = H^{ab}p_ap_b.\]
Then the dimensionless times $\tau$ and $t$ are given by the following scale invariant integrals:
\[  t =   \int \frac{y(xdy - ydx)}{x\sqrt{ x^4+x^3 y-\frac{9}{32}x^2y^2R+\frac{1}{512}y^4S}}, \]

\[ \tau =  \int \frac{(xdy - ydx)}{\sqrt{ x^4 + x^3 y-\frac{9}{32}x^2y^2R+\frac{1}{512}y^4S}}. \]
In the generic case, we find that $u$ varies periodically as a function of the real variable $\tau$ oscillating between two simple real roots of the quartic.  Again if we parametrize the ratio $x:y$ in terms of $\tau$, we may rewrite the integral for $t$ as:
\[ t = \int \frac{yd\tau}{x}.\]
Note that from the point of view of complex integrals, the integral for $t$ always has an additional pole at $x = 0$.   This introduces an ambiguity in the real integral if $S < 0$ and in the imaginary integral if $S > 0$, corresponding to $2\pi $ divided by the modulus of the square root of the quartic at $(x, y) = (0, 1)$.
\eject\noindent
We analyze the behavior of the conformal geometry according to the root structure of the cosmic quartic, where we write the spinor components of $\pi_A$ as $(x, y)$ and the quartic as $H(x, y)$.     For completeness, we include a few cases that wouldnot be realized physically.
\begin{itemize} \item In the (unphysical) case that the quartic is a perfect fourth power, $H(x, y) = (ax + by)^4$, with $a \ne 0$, we have the parametrization: \[ [x, y] = [1 - ab\tau ,  a^2\tau].\] The conformal time $\tau$ ranges over the whole real line.   Then we have:
\[ t =b^{-2}( -ab\tau-\ln(|ab\tau-1|).\]
If $a = 0$, so $H(x, y) = c^2y^4$, with $c > 0$, we have the parametrization: $ [x, y] = [\tau c, 1]$ and then $t = - c^{-1} \ln(|\tau|)$.  Again the conformal time ranges over the whole real line.  In each case,  as $\tau \rightarrow \pm \infty$, the root is approached.  Also, in each case, all ratios $x:y$ with the exception of the triple root are traced out once, as $\tau$ varies.   
\item In the case that the quartic has a triple root  (the two physical possibilities are the origin, $(R, S) = (0, 0)$ and the cusp point,  $(R, S) = (-1, - \frac{27}{8})$),  there are two types of quartics:
\begin{itemize} \item First we may take $H(x, y) = x^3(x + 2ay)$, where $a \ne 0$ (the physical case is $a = \frac{1}{2}$).
Then we may parametrize $(x, y)$ as:
\[ [x, y] = [2a, a^2\tau^2 - 1].\]
Here, a priori, the variable $\tau$ ranges over the whole real line.  Note that $x + 2ay = 2a^3\tau^2$.  As $\tau \rightarrow \pm \infty$, the triple root is approached.  At $\tau = 0$ the simple root is reached, but not crossed.\\We need $\frac{x}{y}$ to be positive, so if $a > 0$, we need $a^2\tau^2 > 1$ and if $a < 0$, we need $a^2\tau^2 < 1$. 
Then we have:
\[ t =   \frac{\tau}{6a}(a^2\tau^2-3).\]
The physical case has $a > 0$ and $\frac{y}{x} > 0$, so occupies the interval $\tau > \frac{1}{a}$ and the age of the universe is infinite.  
\item Second, we may write the quartic as:  $H(x, y) =   (x-3ay)^3(x+ ay)$, where $a \ne 0$ (the physical case is $a = - \frac{1}{8}$).
Then we may parametrize $(x, y)$ as:
\[ [x, y] = [a(12a^2\tau^2+ 1), 4a^2\tau^2 - 1].\]
Here, a priori, $\tau$ ranges over the whole real line. 
We have $x - 3ay = 4a$ and $x + ay = 16a^3\tau^2$.  As $\tau \rightarrow \pm \infty$, the triple root is approached.  At $\tau = 0$ the simple root is reached, but not crossed.\\
We then have:
\[ t = \frac{1}{9a^2}(3a\tau -2\sqrt{3} \arctan(2a\sqrt{3}\tau)).\]
For the physical case that $a < 0$, $\tau$ may be taken to range between  $\frac{1}{2a}$ and $-\frac{1}{2a}$, so the physical age $T$ of the universe is finite and is (in dimensionless units):  \[ T = \frac{1}{9a^2} (4\pi\sqrt{3}  - 9).\] 

 In particular for the case $a = - \frac{1}{8}$, so $(R, S) = (-1, - \frac{27}{8})$,  we have $T =   \frac{256\pi\sqrt{3}}{9} - 64$.
\end{itemize}  Thus, in each case, the ratio $x:y$ for the conformal geometry moves  from the triple root as $\tau \rightarrow - \infty$ to the single root at $\tau = 0$ and then retraces its steps, going back to the triple root as $\tau \rightarrow \infty$.  However the evolution in physical time $t$ occupies only a part of this full trajectory.
\item In the case that the quartic is a perfect square,  with real roots,  (the physical case is $(R, S)  = (-\frac{8}{9}, 0)$), we may write the quartic as follows:
\[ H(x, y) = x^2(x + ay)^2,\hspace{4pt} \textrm{where}\hspace{4pt}  a \ne 0.\]
The physical case is $a = \frac{1}{2}$.
Then we may parametrize $(x, y)$ as:
\[ (x, y) = [ a, e^{a\tau } - 1].\] 
We have $x + ay = a e^{a\tau}$.
 Then, as  $\tau \rightarrow -\infty$, one of the roots is approached and as $\tau \rightarrow \infty$, the other is approached.  Note that in this case, there is a natural periodicity in imaginary $\tau$, with period $2\pi i a^{-1}$.  In particular for the physical case that $H(x, y) = x^2(x + \frac{y}{2})^2$, we have period $4\pi i $.  Physically, if we take $a > 0$, we need $u > 0$, so $\tau > 0$.    Then we have:
\[ t = a^{-2} (e^{a\tau} - a\tau).\]  So $t$ has infinite range.  
\eject\noindent
\item Alternatively, again for the case that the quartic is a perfect square, we may write the quartic in a way that allows the roots to be complex:
 \[ H(x, y) = (x^2 - \epsilon c^2y^2)^2, \hspace{4pt} \textrm{where} \hspace{4pt} c >  0, \epsilon^2 = 1.\]
\begin{itemize} \item If $\epsilon = 1$, then we may parametrize as follows:
\[ [x, y] = [c\cosh(c\tau), \sinh(c\tau)].\]
We have $x + cy = ce^{c\tau}$ and $x - cy =  ce^{-c\tau}$.
 Then, as  $\tau \rightarrow -\infty$, one of the roots is approached and as $\tau \rightarrow \infty$, the other is approached.  Note that in this case, there is a natural periodicity in imaginary $\tau$, with period $\pi i c^{-1}$.  This case is equivalent to the previous case, modulo an appropriate $\textrm{SL}(2, \mathbb{R})$ transformation.  Under that transformation, we find that  $c = \frac{|a|}{2}$, so the periods are in agreement.  Also we have $t = c^{-1} \ln(\cosh(c\tau))$.  Here $\tau > 0$, so the universe has infinite age.
 \item If $\epsilon = -1$, then we may parametrize as follows:
\[ [x, y] = [c\cos(c\tau), \sin(c\tau)].\]
Then the conformal geometry is periodic, with period  $\pi c^{-1}$, covering all ratios $x:y$ in the period.  This case is not physical.
 \end{itemize} 
\item In the case that the quartic has a single repeated root, generically it takes the form:   
\[ H(x, y) = \frac{1}{b}(x+2ay)^2(bx-(b - 1)ay)(x+(b- 1)ay).\]
Here $a \ne 0$ and $b \ne 0$.  Also, without loss of generality, we may take $b$ to be real, with $0 < |b| < 1$ and $b \ne \frac{1}{3}$, or $b$ to be complex with $|b| = 1$ and $\Im(b) > 0$.
Put $c =  |a|\sqrt{|b^{-1}(3 - b)(3b - 1)|} > 0$.
\begin{itemize}\item 
We first assume that $b$ is real and $0 < b < \frac{1}{3}$.\\We parametrize $(x, y)$ as follows:
\[ [x, y]  = [-2a(b-1)(2(1-b)+z(b+1)), 1-6b+b^2+z(b^2 - 1)].\]
Then the equation for $d\tau$ reduces to:
\[ c^2(d\tau)^2 = \frac{dz^2}{1-z^2}, \hspace{10pt} c = |a|\sqrt{|b^{-1}(3 - b)(1 - 3b)|} > 0.\]
The left-hand side is positive, so necessarily $ |z| < 1$.  
Then we may take $z = \epsilon \sin(c\tau)$, where $\epsilon^2 = 1$, giving:
\[ [x, y]  = [-2a(b-1)(2(1-b)+ \epsilon\sin(c\tau)(b+1)), 1-6b+b^2+\epsilon\sin(c\tau)(b^2 - 1)].\]
When $\sin(c\tau) = \epsilon$ we are at the simple root $2(b- 3) [a(b -1), b]$.\\
When $\sin(c\tau) = -\epsilon$ we are at the simple root $ 2(1 - 3b) [a(1 - b), 1]$.\\
So the conformal metric oscillates forever between the two roots, never crossing either, and avoiding the double root completely (since $x + 2ay = 2a(1 - 3b)(3 - b) \ne 0$).  Its period is $2\pi c^{-1}$.  For the physical metrics we may take $a = \frac{b}{(b +1)^2}$.    Then we have $z$ in the interval $(-1, b_0)$ where we have:
\[  b_0 = \frac{b^2 - 6b + 1}{1 - b^2}.\]
Putting $z = \frac{ 1- p^2}{1 + p^2}$, the time function is then:
\[ t = - \frac{\arctan(p)(b + 1)^4}{\sqrt{b^3(b - 3)(3b - 1)}}+(b + 1)^4 b^{-\frac{3}{2}} (b - 1)^{-1} \arctan\left(\frac{(b - 3)p}{\sqrt{(3b-1)(b-3)})}\right).\]
The time $T$ elapsed during one cycle of the universe, going from the big bang to its maximum radius and back again to zero radius is:
\[    2b^{-\frac{3}{2}} (b + 1)^4\left(( 1- b)^{-1}\arctan(b^{-\frac{1}{2}})  +\frac{\arctan\left(\sqrt{\frac{3b-1}{b(b - 3)}}\right)}{\sqrt{(b-3)(3b-1)}}\right).\]
This diverges as $b \rightarrow 0^+$, when the double root becomes a triple root.
\item We next assume that $b$ is real and $ \frac{1}{3}< b < 1$, or $-1 < b < 0$.\\
\\We parametrize $(x, y)$ as follows:
\[ [x, y]  = [-2a(b-1)(4(1-b)z+(z^2 + 1)(b+1)), 2(1-6b+b^2)z+(z^2+1)(b^2 - 1)].\]
Then the equation for $d\tau$ reduces to:
\[ c^2(d\tau)^2 = \frac{dz^2}{z^2 }, \hspace{10pt} c = |a|\sqrt{|b^{-1}(3 - b)(1 - 3b)|} > 0.\]
We may take $z = \epsilon e^{c\tau}$, where $\epsilon^2 = 1$. So we have:
\[ [x, y] = [-2a(b - 1)(4(1 - b) \epsilon e^{c\tau} + (e^{2c\tau} + 1)(b+1)), 2(1 -6b + b^2)\epsilon e^{c\tau}+  (e^{2c\tau} + 1)(b^2 - 1)].\]
Also we have:
\[ x + 2ay = 4az(3b-1)(b - 3), \]
\[ (bx-(b - 1)ay) = -a(b-1)(3b-1)(b+1)(z-1)^2, \]
\[ x+ (b - 1)ay = a(b-1)(b-3)(b+1)(z+1)^2.\]
As $\tau \rightarrow - \infty$, we approach the double root.   At $\tau = 0$ we are at the root $\frac{x}{y} = \frac{(b - 1)a}{b}$ if $\epsilon = 1$, or we are at the root $\frac{x}{y} = (1 - b)a$ if $\epsilon = -1$.   This root is not crossed and as $\tau \rightarrow \infty$, we go back to the double root.   The motion is periodic in imaginary $\tau$ with period $2\pi c^{-1}i $.  Note that the values of the ratio $\frac{x}{y}$ at $\tau$ and $-\tau$ are equal. 
With $a = \frac{b}{(b + 1)^2}$ and $-1 < b < 0$, the physical  time function is:
\[ t = \frac{(b + 1)^2}{2bc(b - 1)}\left((1 - b)\ln(|z|) +2 \textrm{arctanh}\left(\frac{(2(1-b)+z(1 + b))}{\sqrt{(3b-1)(b-3)}}\right)\sqrt{(3b-1)(b-3)}\right).\]
With $a = \frac{b}{(b + 1)^2}$ and $\frac{1}{3} < b < 1$, the physical  time function is:
\[ t = \frac{(b + 1)^2}{2bc(b - 1)}\left((1 - b)\ln(|z|) +2 \textrm{arctan}\left(\frac{(2(1-b)+z(1 + b))}{\sqrt{-(3b-1)(b-3)}}\right)\sqrt{-(3b-1)(b-3)}\right).\]
In either case, the universe has infinite age, since $t$ diverges as $z \rightarrow 0$, so as the double root is approached.
\item Finally we assume that $b$ is complex with $|b| = 1$, with $\Im(b) > 0$.  Write $\displaystyle{b = \frac{1 + ip}{1 - ip}}$, where $p $ is real and positive.  Explicitly, we have $\displaystyle{p =  \frac{i(1 - b)}{1 + b}}$. Then make the parametrization:
\[ [x, y] = [-2ap(z^2-4zp-1),  pz^2 + 2z(2p^2+1)-p].\]
With this parametrization, the equation for $d\tau$ becomes:
\[ c^2d\tau^2 =\frac{dz^2}{z^2}, \hspace{10pt} c = |a|\sqrt{|b^{-1}(3 - b)(1 - 3b)|} > 0.\]
So  $z = \epsilon e^{c\tau}$, where $\epsilon^2 = 1$ and $\tau$ ranges over all the reals.
Then we have $x + 2ay =  4az (4p^2 + 1)$.  So the metric cycles once around,  starting and ending at the double root as $\tau \rightarrow \pm \infty$.   As it does so, all values of the ratio $x:y$ are traced once, except for $x:y = - (2a)^{-1}$, the double root.  Note that this metric has a natural period of $2\pi  c^{-1}i $ in imaginary $\tau$.   In this case, the points parametrized by $z$ and $z'$ agree if $zz' = - 1$, so we can also make a "Mobius" identification in the complex, identifying $\tau$ with $- \tau + \pi i c^{-1}$.  The time function is:
\[ t  = -4p^{-1}(1 + p^2)^{-\frac{3}{2}} (1 + 4p^2)^{-\frac{1}{2}}\left(p\ln(|z|)-2(1 + 4p^2)^{\frac{1}{2}}\textrm{arctanh}\left(\frac{z-2p}{\sqrt{4p^2+1}}\right)\right).\]
Again the physical age of the universe is infinite.
\end{itemize} 
For these cases, the corresponding values of the parameters $R$ and $S$ are:
\[ [R, S] = - \left[\frac{32 b(3-2b+3b^2)}{9(b+1)^4}, \frac{2048b^3(b-1)^2}{(b+1)^8}\right].\]
\begin{itemize} \item When $b \rightarrow \frac{1}{3}$ we go to the cusp point $(-1, -\frac{27}{8})$.  \item When $b \rightarrow 1$, we go to the point $(-\frac{8}{9}, 0)$.   \item When $b \rightarrow 0$, we go to the point $(0, 0)$.  \item  The $b$-interval  $(\frac{1}{3}, 1)$ traces out the boundary curve IV-VA in the third quadrant.  The period goes steadily decreases as $R$ increases, going from $\infty$ at the cusp point, to $4\pi i$ at the point $(-\frac{8}{9}, 0)$.
 \item  The $b$-interval  $(0 , \frac{1}{3})$ traces out the curve IV-VB in the third quadrant.
 \item  The $b$-interval  $(-1 , 0)$ traces out the curve I-II in the first quadrant.  The period decreases from $\infty$ as $R \rightarrow 0^+$ to zero as $R\rightarrow \infty$. 
\item When $b$ is complex, not real and $|b| = 1$, we trace out the boundary curve II-III boundary curve. Put $\theta_0 = \arctan(\sqrt{8})$ and denote the argument of $b$ by $\theta$, so $0 < \theta < \pi$.   The $\theta$-interval $(0, \theta_0)$, or $0 < p < \frac{1}{\sqrt{2}}$ traces out the  then we are in the part of the II-III boundary curve in the second quadrant.  The $\theta$-interval $( \theta_0, \pi)$, or $p > \frac{1}{\sqrt{2}}$ traces out the part of the II-III boundary curve in the first quadrant.  When $\theta = \theta_0$, or $p = \frac{1}{\sqrt{2}}$ we are at the II-III-boundary point on the $S$-axis  at $(0, 54)$.   The period steadily decreases as $R$ increases, starting at $4\pi i$ at $(- \frac{8}{9}, 0)$, going down to $\frac{4}{3} \pi i \sqrt{2} $ at $(0, 54)$ and going towards zero as $R\rightarrow \infty$. \end{itemize} 
In summary, along the upper curve $S = S_+(R)$, except where it meets the lower curve $S = S_-(R)$ at the cusp point, and along the part of the lower curve $S = S_-(R)$ in the first quadrant, the metrics have a periodicity in imaginary $\tau$.   In each case the period is $\displaystyle{\frac{2\pi i |b +1|^2}{\sqrt{|b(3b - 1)(b - 3)|}}}$.  Note that this goes to infinity as $b \rightarrow \frac{1}{3}$, which gives the cusp point $(R, S)  = (-1, -\frac{27}{8})$,and as $b \rightarrow 0$, which gives the origin.  Also as $b \rightarrow 1$, the period goes to $4\pi i $.   This is agrees with the imaginary period found previously, for this case, the case $(R, S)  = (-\frac{8}{9}, 0)$.
\item The non-generic cases with two repeated roots are the cases with $S = 0$.   For these, we may write the quartic $H(x, y)$ as follows:
\[ H(x, y) = x^2(x^2 - 2bxy - \epsilon cy^2).\]
Here $b$ and $c$ are real, with $c > 0$ and $b^2 + \epsilon c \ne 0$.  Note that $R$ and $\epsilon $ have the same sign.
\begin{itemize} \item When $\epsilon = - 1$, (corresponding to $R < 0$, which gives the II-V boundary curve for $c < b^2$ and the III-IV boundary curve for $c > b^2$),  we make the parametrization:
\[ [x, y] = [4cz,  4(b^2-c) + z^2 + 4bz].\]
Note that $x^2 - 2bxy + cy^2 = c(z^2-4b^2+4c)^2$.
Then we have $c d\tau^2 = z^{-2} dz^2$, so we may take $z  = \alpha e^{\tau\sqrt{c}}$, where $\alpha^2 = 1$.  When $\tau \rightarrow \pm \infty$, the ratio $x:y$ goes to the double root.  If $b^ - c < 0$, it simply goes through all values of the ratio $x:y$, except for $0$, as $\tau$ ranges through the real numbers.  If, instead, $b^2 - c > 0$, then $x:y$ goes to the root: $2c: \alpha \sqrt{b^2-c}  + 1$, when $e^{2\tau\sqrt{c}} = 4 (b^2 - c)$.  So the solution goes from the double root, to a single root, which it does not cross, and then back to the single root.  In both cases, the solution is periodic in imaginary $\tau$ with period $2\pi i c^{-\frac{1}{2}}$.   In terms of the physical variable $R$, the period is  $\frac{16\pi i}{3 \sqrt{-2R}}$.  In particular the period goes to infinity as $R\rightarrow 0^-$ and the period goes to $4\pi i$ as $R \rightarrow - \frac{8}{9}$, both as expected.  As $R\rightarrow - \infty$, the period goes to zero.  The physical time function is:
\[ t = -\frac{1}{4zc^{\frac{3}{2}}}(-z^2-4bz\ln(|z|)+4b^2-4c).\]
The physical age of the universe is infinite.
\item When $\epsilon = 1$, (corresponding to $R > 0$, so the I-IV boundary curve),we make the parametrization:
\[  [x, y] =  [- c ,  b + z\sqrt{c+b^2}].\]
Note that $x^2 - 2bxy - cy^2 = c(1-z^2)(c+b^2)$.
Then we have $\displaystyle{c d\tau^2 =  \frac{dz^2}{1 - z^2}}$, so we may take $z = \cos(\tau\sqrt{c})$.   The solution is periodic in $\tau$ with period $\tau_0 = \frac{2\pi}{\sqrt{c}} = \frac{16\pi }{3 \sqrt{-2R}}$.   For $n$ any integer, at $\tau = 2n \tau_0$, the solution is at a single root, then goes to the double root when $\cos(\tau\sqrt{c}) = -\frac{b}{\sqrt{c + b^2}}$, which it passes through, then goes to the other single root at  $\tau = (2n + 1)\pi \tau_0$, then returns to the other single root (after passing through the double root again), at $\tau = (2n + 2)\pi \tau_0$.  Neither single root is crossed.   The physical time function is;
\[ t = c^{-\frac{3}{2}}(b\tau \sqrt{c}+\sqrt{c+b^2}\sin(\tau \sqrt{c})).\]
If we start the universe at a big bang with $u = \infty$, we need $y = 0$, so $z = - \frac{b}{\sqrt{c + b^2}}$.  Then we return to a big crunch the next time that $z$ returns to its initial value.   During that time the quantity $\tau \sqrt{c}$ increases by $2\pi$, so the full cycle of the universe lasts the finite time $2\pi |b| c^{-\frac{3}{2}}$.   The physical value of $b$ is $-\frac{1}{2}$,  giving the lifetime as $\pi c^{-\frac{3}{2}} = \frac{512\pi}{27} (-2R)^{-\frac{3}{2}}$.  
\end{itemize} 
\end{itemize} 
\eject\noindent
\section{The generic quartics: the two periods}
Lastly we need to consider the generic case, where the roots are all distinct.    We may then write the quartic at first in the following form:
\[ H(x, y) = ((x - ay)^2 + by^2))((x - cy)^2 + dy^2).\]
Here $a$, $b$, $c$ and $d$ are real and the roots are $\frac{x}{y} = a \pm \sqrt{-b}, c \pm \sqrt{-d}$ and $b$ and $d$ must be non-zero.  Multiplying out, we find that for the physical quartics, we may take:
\[ c = - \frac{1}{2}(1 + 2a), b = \frac{1}{4}a(1 - p), d = \frac{1}{4} c( 1+ p) = - \frac{1}{8} (1 + 2a)( 1+ p).\]
Here $a \ne 0$,  $a\ne -\frac{1}{2}$, $p \ne 1$ and $p \ne -1$.   Also we have: 
\[ R =  \frac{4}{9} (p(1 + 4a)+ 2a^2+8a-1),  S = 16a(1+2a)(1 - p +4a)^2.\]
If $p = 1 + 4a$, then $H(x, y)$ has a factor of $x^2$, contrary to hypothesis.  So we have $p \ne 1+ 4a$.  So $\Lambda$ is non-zero and has the sign of $a(1 + 2a)$. Then we find that the four roots are always distinct, except for the following value of $p$, which must be ruled out :
\[ p = (4a + 1)(64a^2 + 32a + 1).\]
This parametrization is symmetrical under $(a, p) \rightarrow (- a - \frac{1}{2}, - p)$, so, without loss of generality, we may henceforth assume that $a \ge - \frac{1}{4}$.   In particular $2a + 1 > 0$ and $S$ and $\Lambda$ have the sign of $a$.
\begin{itemize} \item The four roots are real if and only if $a(p - 1)$ and $(1 + 2a)(p + 1)$ are both positive, so if and only if $a > 0$ and $p > 1$ (so $\Lambda >0$) , or $- \frac{1}{4} \le  a < 0$ and $- 1 < p < 1$ (so $\Lambda < 0$).
\item The four roots are non-real if and only if the $a(p - 1)$ and $ (1 + 2a)( p+ 1)$ are both negative, so if and only if $a > 0$ and $p < -1$.  In this case $\Lambda > 0$.
\item Exactly two roots are real if and only if  $a(2a +1)(p^2 - 1)< 0$ is negative, so if and only if $a > 0$ and $- 1 < p < 1$ (so $\Lambda > 0$), or $ - \frac{1}{4} \le a < 0$ and $|p| > 1$ (so $\Lambda < 0$).
\end{itemize} 
\eject\noindent
We change notation to analyze each of the three cases in turn.
\begin{itemize} \item If none of the roots is real, then by an appropriate $\textrm{GL}(2, \mathbb{R})$ transformation, we can put the quartic in the following standard form:
\[ H(x, y) = A^2(x^2 +y^2)(x^2 + k^2 y^2).\]
Here $0 < k < 1$ and $A > 0$.    Put $k' = \sqrt{1 - k^2}$, so $0 < k' < 1$ also.
We may construct the required transformation, as follows, where we assume that the roots are known, for example, using the parametrization given above.  Let $H(x, y)$ be given initially as follows:
\[ H(x, y) =  ((x- ya)^2 + b^2y^2)((x - yc)^2 + d^2y^2).\]
Here $a$, $b$, $c$ and $d$ are real and $b$, $d$ are positive. Also the four complex roots,  $a \pm ib$ and $c \pm id$ are assumed to be pairwise distinct. Then we make the parametrization:
\[ [x, y] = [s(a^2 + b^2 - bkd - ac)+akd - bc, \hspace{4pt} s(a - c) + kd-b].\]
The inverse transformation is:
\[ s = \frac{x(b-kd)+y(akd-bc)}{x(a-c)- y(a^2+b^2-bkd-ac)}.\]
Here the real parameter $k$ is required to obey the quadratic equation:
\[ bd(k^2+1)-(d^2+c^2-2ca+a^2+b^2)k = 0.\]
This quadratic equation for the unknown $k$ has discriminant: 
\[ ( ((a+ ib)-(c+id))((a-ib)-(c- id))((a+ib)-(c- id)))((a - ib)-(c+ id))).\]
This is always positive, so the two roots for $k$ are real, distinct, positive and have product $1$, so there is exactly one root for $k$ in the interval $(0, 1)$, which gives us the required value of $k$. 
\eject\noindent
Then the parametrization is always well-defined and with respect to this parametrization, the equation for $\tau$ becomes:
\[ d\tau^2 = \frac{k ds^2}{bd(1 + s^2)(k^2 + s^2)}.\]
So we have $A = \sqrt{ k^{-1}bd}$.
Then we have
\[ Ak\tau = \int \frac{k}{\sqrt{(1 + s^2)(k^2 + s^2)}} ds = - i \textrm{EllipticF}(is, k^{-1}).\]
For fixed $k$, the inverse function is:
\[ s = - i \textrm{JacobiSN}(iAk\tau, \frac{1}{k}).\]
As $\tau$ varies over the real line, the ratio $x:y$ runs through all possible values.   Then the complete integral gives the conformal period $T$ of the universe: the amount that $\tau$ elapses for the universe to return to its original state.   We have:
\[ AkT =  k\int_{-\infty}^\infty \frac{1}{\sqrt{(1 + s^2)(k^2 + s^2)} }ds  = 2k\textrm{EllipticK}(k').\]
As an analytic function of $\tau$, $s$ is doubly periodic.   The second period is imaginary: $iT'$, for $T'$ real and is given by $\sqrt{-1}$ times twice the integral the integral of $\frac{1}{\sqrt{H}}$ along the imaginary axis from one root to another.   So here we have:
\[ AkT' =  2k\int_{-k}^k \frac{1}{\sqrt{(1- s^2)(k^2- s^2)}} ds = 4k\textrm{EllipticK}(k).\] 
 Since as $\tau$ varies, all ratios $y:x$ are traced out, the integral for the physical time function has a pole when $x = 0$ and the physical age of the universe is infinite.  These universes all have $\Lambda > 0$.
 \eject\noindent  
\item If exactly two of the roots are real, then by an appropriate $\textrm{GL}(2, \mathbb{R})$ transformation, we can put the quartic in the following standard form:
\[ H(x, y) = A^2(y^2 - x^2)(x^2 + k^2 y^2).\]
Here $k > 0$ and $A > 0$. 
We may construct the required transformation, as follows, where we assume that the roots are known, for example using the parametrization of the last section.  Let $H(x, y)$ be given initially as follows:
\[ H(x, y) =  ((x- ya)^2 - b^2y^2)((x - yc)^2 + d^2y^2).\]
Here $a$, $b$, $c$ and $d$ are real and $b$, $d$ are positive.  Then the four roots,  $a \pm b$ and $c \pm id$ are pairwise distinct.  Then we make the parametrization:
\[ [x, y] = [s(a^2-b^2 + bdk - ac) + akd-cb, (a - c)s + kd-b.\]
The inverse transformation is:
\[ s = \frac{x(kd-b)+y(bc-akd)}{x(c-a)+y(a^2-b^2+bkd-ac)}.\]
Here the parameter $k$ is required to obey the quadratic equation:
\[ bd(k^2 - 1)- (b^2-d^2-c^2-a^2+2ac)k = 0.\]
The product of the roots is $-1$, so both roots are real and exactly one is positive, which gives us the required value of $k$.  Then the parametrization is always well-defined and with respect to this parametrization, the equation for $\tau$ becomes:
\[ d\tau^2 = \frac{k ds^2}{bd(1 - s^2)(s^2 + k^2)}.\]
So necessarily $|s| < 1$ and we have $A = \sqrt{bdk^{-1}}$.
The solution can be taken to be:
\[ Ak\tau =  \textrm{EllipticF}(s, ik^{-1}) =  \frac{k\textrm{EllipticK}(\frac{1}{\sqrt{1 + k^2}}) - k\textrm{EllipticF}(\sqrt{1 - s^2},\frac{1}{\sqrt{1 + k^2}})}{\sqrt{ 1+ k^2}}.\]
The inverse solution is:
\[ s = - i \textrm{JacobiSN}(Ak\tau, ik^{-1}).\]
As $\tau$ varies, the solution oscillates smoothly between the value $s = -1$, when we are at the root  $x:y = (a -b):1$  and the value $s = 1$, when we are at the root $x:y = (a + b):1$.
The complete integral gives the conformal period $T$ of the universe: the amount that $\tau$ elapses for the universe to return to its original state.   We have:
\[AkT =  2k \int_{-1}^1 \frac{ds}{\sqrt{(1 - s^2)(s^2 + k^2)}} = \frac{4k}{\sqrt{1 + k^2}} \textrm{EllipticK}(\frac{1}{\sqrt{1 + k^2}}) = 4\textrm{EllipticK}(ik^{-1}).\]
 The second period is imaginary $iT'$, for $T'$ real and is given as follows:  
 \[ AkT' =  2k \int_{-k}^k \frac{ds}{\sqrt{(1 + s^2)(k^2 - s^2)}}  = \frac{4k}{\sqrt{1 + k^2}} \textrm{EllipticK}(\frac{k}{\sqrt{1 + k^2}}).\]
 In the case that $\Lambda > 0$, the evolution goes through $x =0$, so the integral for the physical time diverges to infinity there and the universe has infinite age.    In the case that $\Lambda < 0$, the evolution avoids $x =0$, so the integral for the physical time is finite and the universe has finite age.   
\item If all four of the roots are real, then by an appropriate $\textrm{GL}(2, \mathbb{R})$ transformation, we can put the quartic in the following standard form:
\[ H(x, y) = A^2(y^2 - x^2)(x^2 - k^2 y^2).\]
Here $k > 0$ and $A > 0$. 
We may construct the required transformation, as follows, where we assume that the roots are known, for example using the parametrization of the last section.  Let $H(x, y)$ be given initially as follows:
\[ H(x, y) =  ((x- ya)^2 - b^2y^2)((x - yc)^2 - d^2y^2).\]
Here $a$, $b$, $c$ and $d$ are real and $b$, $d$ are positive.  Also the four roots,  $a \pm b$ and $c \pm d$ are assumed to be pairwise distinct.  Finally, without loss of generality, we may assume that both the roots $a \pm b$ are larger than both the roots $c \pm d$, so we have: $a + b > a - b > c + d > c - d$. 
\eject\noindent
Then we make the parametrization:
\[ [x, y] = [(a^2-b^2- bdk-ac)s-akd - bc, (a-c)s-b-dk].\]
The inverse transformation is:
\[ s   = \frac{x(b+kd)-y(akd+bc)}{x(a-c)+y(-a^2+b^2+bkd+ac)}.\]
Here the parameter $k$ is required to obey the quadratic equation:
\[f(k) =  bd(k^2 + 1)+ (b^2+d^2-c^2-a^2+2ac)k = 0.\]
The product of the roots is $1$.  Also we have:
\[ f(1) = - ((a- b)-(c+d))((a+b)-(c-d)) < 0.\]
So the roots are real and there is exactly one root with $k > 1$, which gives us the required value for $k$.
Then the parametrization is always well-defined and with respect to this parametrization, the equation for $\tau$ becomes:
\[ d\tau^2 = \frac{k ds^2}{bd( s^2 - 1)(k^2 - s^2)}.\]
So necessarily $1 < |s| < k$ and we have $A = \sqrt{bdk^{-1}}$.
Then we may take:
\[ Ak\tau = \textrm{EllipticF}(\sqrt{\frac{1-t^{-2}}{1-k^{-2}}}, \sqrt{1-k^{-2}}) - \textrm{EllipticK}(\sqrt{1 - k^{-2}}).\]
The inverse expression is:
\[ s^{-2}  = 1 +(k^{-2}- 1)\textrm{JacobiSN}(ak(\tau+ \tau_0), \sqrt{1-k^{-2}})^2, \]
\[ \tau_0 =  (ak)^{-1}\textrm{EllipticK}(\sqrt{1 - k^{-2}}).\]
The real period $T$ is given by the formula:
\[ AkT = 2k \int_1^k \frac{ds}{\sqrt{ (s^2 - 1)(k^2 - s^2)}} = 2\textrm{EllipticK}(\sqrt{1 - k^{-2}}).\]
The imaginary period is $iT'$, where we have:
\[ AkT' = k \int_{-1}^1  \frac{dt}{\sqrt{ (1 - s^2)(k^2 - s^2)}} = 2\textrm{EllipticK}(k^{-1}). \]
\eject\noindent
There are two possible evolutions, according to whether or not $s > 0$. 
\begin{itemize} \item  If $s > 0$, as $\tau$ varies $s$ cycles smoothly between its minimum value of $1$ and its maximum value, $k >1$.  At $s = 1$, we are at the root $x:y = (a + b):1$; at $s = k$, we are at the root $x:y = (c -d):1$.  Neither root is crossed during the evolution and the other two roots are avoided completely.
 \item  If $s < 0$, as $\tau$ varies $s$ cycles smoothly between its minimum value of $-k$ and its maximum value, $-1$. At $s = -1$, we are at the root $x:y = (a - b):1$; at $s = -k$, we are at the root $x:y = (c +d):1$.  Neither root is crossed during the evolution and the other two roots are avoided completely.
\end{itemize} 
 In the $\Lambda < 0$ case the universe has finite lifetime.    In the $\Lambda > 0$ case, the universe has infinite lifetime.

\end{itemize} 
Finally note a general fact that has emerged in these case by case analyses: that near a simple zero of the cosmic quartic, the conformal time parameter is related to a spinor variable $u$ which vanishes at that zero, by a formula of the form $u = c(\tau - \tau_0)^2$, to lowest order, with $c \ne 0$.  Then the metric is smooth at the zero, in terms of the conformal time parameter $\tau$ and the zero \emph{is not crossed}.  Thus, in the generic case that the quartic has no repeated roots, the conformal geometry must  oscillate perpetually between two adjacent real zeroes of the cosmic quartic, if the cosmic quartic has at least one such zero (since then it must have at least two such).   If the quartic has no real zeros, then the conformal time is unbounded and the spinor parameter $\pi_A$ ratios range over the full projective circle.  In this case we may regard the conformal geometry as being periodic in real conformal time, with period given by the integral of $y^{AB} \pi_{A} d\pi_B$ over the projective circle.  If the quartic has a repeated real root, then this root can be crossed.
\eject\noindent
\section{Summary and discussion}
We have shown that cosmological metrics can be put in the following standard form, using a real projective spinor co-ordinate $\pi_A$, where $\pi_A$ is equivalent to $k\pi_A$ for any non-zero real $k$:
\[ ds^2 = u^{-2} (d\tau^2 - g_\kappa), \]
\[ d\tau = y^{AB} \pi_A d\pi_{B},  \hspace{10pt} y^{AB} = - y^{BA}, \]
\[  y^{AB}y^{CD} \pi_{A}\pi_{C}H_{BD}^{EF}\pi_E\pi_F = 4.\]
\[ u = \frac{\alpha^A \pi_A}{\beta^B\pi_B}.\]
Here $g_\kappa$ is a three-geometry of constant curvature $\kappa$, with $\kappa^3 = \kappa$.  Also $H_{AB}^{CD}$ is a constant spinor, which is symmetric in the pairs $AB$ and $CD$ and is trace-free.  Also $\alpha^A$ and $\beta^A$ are fixed (linearly independent) spinors. The co-ordinate $\tau$ measures conformal time:  the vector field $\partial_\tau$ is the conformal Killing vector orthogonal to the surfaces of homogeneity.  The physical time variable $t$ is given in terms of $\tau$ by the differential relation:
\[ dt = u^{-1} d\tau.\] 
The equation $y^{AB}y^{CD} \pi_{A}\pi_{C}H_{BD}^{EF}\pi_E\pi_F = 4$, with $\pi_A$ complex, defines a torus in the case that $H_{AB}^{CD}$ is generic.  An equivalent but perhaps slightly less elegant formulation of the metric is:
\[ ds^2 = U^{-1}\left(\frac{g^{ab} dp_a dp_b}{2H^{cd}p_cp_d} - G_\kappa\right).\]
Here $g^{ab}$ is a flat metric of signature $(2, 1)$ in three-dimensions and $H_a^b$ is a constant tensor, with $H^{ab} = g^{ac} H_c^b$ symmetric.  Also the metric is restricted to the null cone in $p_a$ space: $g^{ab}p_ap_b = 0$ and $p_a$ is identified with $sp_a$, for $s > 0$ any real number.  Also $\displaystyle{U = \frac{p_au^a}{p_bv^b}}$, for constant null vectors $u^a$ and $v^a$ (here the connection with the previous metric is given by using  writing $p_a = \pi_A \pi_B$, $g^{ab} = \epsilon^{A(C} \epsilon^{D)B}$ (where $\epsilon^{AB} = - \epsilon^{BA}$),  $u^a = \alpha^A\alpha^B$ and $v^a = \beta^A\beta^B$.   From this viewpoint there are obvious questions and possible generalizations: why does $U$ take this form? Why is $H_a^b$ constant? Is the six-dimensional metric of signature $(2, 4)$ given by $ds^2$, but with the constraint that $p_a$ be null removed, of some physical interest?  Is $H_a^b$ determined from structure at the big bang?  At the time of writing we do not know the answer to these questions.  
\eject\noindent However it does seem not unlikely that at the big bang, there is some kind of phase transition for which the tensor $H_a^b$ as an order parameter: in other words that it's specific value is not a priori predictable, but it belongs to a class of such parameters that might be predictable as a whole.  Then once the tensor $H_a^b$ is set, the basic disposition of matter, radiation, topology and cosmological constant is imprinted for ever on the universe.\\\\
In the generic case we have found that the conformal geometry simply oscillates for ever between two roots of the quartic spinor polynomial defined by $H_{AB}^{CD}$ (unless there are no real roots, in which case all values of the projective spinor are visited repeatedly).   This may be relevant for theories of ekpyrotic [7]  or cyclic universes [8].   The physical geometry then occupies a part of the conformal evolution.\\\\
Next we have observed that generically, in the complex, the differential form defining the time $dt = u^{-1}d\tau$ has an imaginary period, corresponding to the integral along the path in imaginary $\tau$ going once around the torus defined by $H_a^b$, with the interval of integration in $\tau$ given by a complete elliptic integral.   We have shown how to organize the appropriate co-ordinate transformations to put the  period integral into a standard form used in elliptic integral theory.  Given $H_a^b$, it is possible to compute the period integral in terms of elliptic functions.  This imaginary period is unambiguous in the case that $\Lambda < 0$.  However in the case that $\Lambda > 0$, there is an ambiguity that arises from the contribution of the pole at $u = 0$.  One could eliminate the ambiguity if one simply prescribes that the contour is not allowed to stray into the the pre-big-bang era: $\Re(u) < 0$.  However it is not clear whether this is physically the right thing to do.  So at least in the case that $\Lambda < 0$, we have a well-defined periodicity in imaginary time, which is directly calculable in terms of the ingredients of the cosmic balance equation.     Now since the work of Hawking on black hole radiation, it has been realized that periodicity in imaginary time is associated with a mass and via Planck's constant with a temperature.  Using standard values for the fundamental constants of nature, the formula for the mass is   $3.2126(10^{34})(\Im{t})$ kilograms, where the periodicity $\Im(t)$ in imaginary time is measured in seconds.  So we have a natural mass scale for such universes, in principle measurable from cosmic data.   
\eject\noindent
One would like to give an actual value to the parameters $R$ and $S$ that determine our tensor $H_a^b$.   In section three above these parameters are given directly in terms of the quantities involved in the cosmic balance equation.   However their values are very sensitive to the value of the radiation component of the universe, which seems to have been neglected in current experimental measurements, because it is believed to be small.   Instead these measurements have focussed on the matter component and on the $\Lambda$ term.   It should perhaps be emphasized that Einstein's equations are inexorable and apply to the entire evolution of the universe.  Therefore, what is needed, it seems is a global fit for these parameters, taking into account the whole history of the universe.   Even to determine these parameters with an error of less than ten percent would be a terrific achievement.  Note that there is one combination of these parameters namely $SR^{-3}$ which is independent of the radiation density, so this combination may be easier to pin down. \\\\
Finally it is worth mentioning that conformally flat cosmological models often have a nice description in terms of twistor theory [6].  In the present case the three-parameter family of world lines of the vector field $\partial_\tau$ can be described by a three-parameter family of quadrics in projective twistor space.  Using standard twistor variables and two-component spinors, these quadrics have the equation:
\[ \omega^A t_a \pi^{A'} + x^{A'B'}\pi_{A'} \pi_{B'} = 0.\]   Here the twistor is $Z^\alpha = (\omega^A, \pi_{A'})$, the spinor $x^{A'B'} = x^{B'A'}$ represents the particular world-line and the timelike vector $t_a$ represents its tangent vector to the world-line.  After deleting the line at infinity, each quadric (of real dimension four) has topology $\mathbb{S}^2 \times \mathbb{C}$.  Then the identification in imaginary time corresponds to an identification in twistor space of the form $(\omega^A, \pi_{A'}) \equiv (\omega^A + st^a\pi_{A'} , \pi_{A'})$, with $s$ real.  This wraps up the $\mathbb{C}$-factor giving the quadrics the topology of the product of a sphere and a cylinder.  If we further identify corresponding to a real periodicity in the variable $\tau$, then the cylinder factor is wrapped into a torus.
\eject
\end{document}